\documentclass[useAMS,usenatbib]{mn2e}

\usepackage{graphicx}
\usepackage{amsmath}
\usepackage{amssymb}
\usepackage{hyperref}

\title[Cosmic rays and gamma rays from supernova remnants]{Acceleration of cosmic rays and gamma--ray emission from supernova remnants in the Galaxy}
\author[P. Cristofari et al.]
{P. Cristofari$^{1}$\thanks{E-mail: pierre.cristofari@apc.univ-paris7.fr},
S. Gabici$^{1}$,
S. Casanova$^{1,2,3}$,
R. Terrier$^{1}$,
and E. Parizot$^{1}$
\\
$^{1}$APC, AstroParticule et Cosmologie, Universit\'e Paris Diderot, CNRS/IN2P3, CEA/Irfu, Observatoire de Paris, Sorbonne Paris Cit\'e,\\ 10, rue Alice Domon et L\'eonie Duquet, 75205 Paris Cedex 13, France \\
$^{2}$Max-Planck-Institut f\"ur Kernphysik, Saupfercheckweg 1, 69117 Heidelberg, Germany \\
$^{3}$Unit for Space Physics, North-West University, Potchefstroom 2520, South Africa
}

\begin{document}

\date{}

\pagerange{\pageref{firstpage}--\pageref{lastpage}} \pubyear{}

\maketitle

\label{firstpage}

\begin{abstract}
Galactic cosmic rays are believed to be accelerated at supernova remnant shocks. Though very popular and robust, this conjecture still needs a conclusive proof. The strongest support to this idea is probably the fact that supernova remnants are observed in gamma--rays, which are indeed expected as the result of the hadronic interactions between  the cosmic rays accelerated at the shock and the ambient gas. However, also leptonic processes can, in most cases, explain the observed gamma--ray emission. This implies that the detections in gamma rays do not necessarily mean that supernova remnants accelerate cosmic ray protons.
To overcome this degeneracy, the multi--wavelength emission (from radio to gamma rays) from individual supernova remnants has been studied and in a few cases it has been possible to ascribe the gamma--ray emission to one of the two processes (hadronic or leptonic). 
Here we adopt a different approach and, instead of a case--by--case study we aim for a population study and we compute the number of supernova remnants which are expected to be seen in TeV gamma rays above a given flux under the assumption that these objects indeed are the sources of cosmic rays.
The predictions found here match well with current observational results, thus providing a novel consistency check for the supernova remnant paradigm for the origin of galactic cosmic rays.
Moreover, hints are presented for the fact that particle spectra significantly steeper than $E^{-2}$ are produced at supernova remnants. Finally, we expect that several of the supernova remnants detected by H.E.S.S. in the survey of the galactic plane should exhibit a gamma--ray emission dominated by hadronic processes (i.e. neutral pion decay). 
The fraction of the detected remnants for which the leptonic emission dominates over the hadronic one depends on the assumed values of the physical parameters (especially the magnetic field strength at the shock) and can be as high as roughly a half.
\end{abstract}


\section{Introduction}

Though cosmic Rays (CRs) have been discovered about a century ago \citep{deangelis}, the question about their origin is still a matter of discussion \citep[see e.g.][for reviews]{lukereview,etiennereview}. \citet{baade} have been the first to propose that supernovae are the sources of CRs, and this still remains the most popular scenario to explain the origin of galactic CRs. The particle energy that marks the transition between galactic and extra--galactic CRs is believed to be located between the knee and the ankle which are observed in the CR spectrum at particle energies of $\approx 4$~PeV and a few $10^{18}$~eV, respectively. Thus, the acceleration mechanism connected to supernovae must be able to accelerate particles up to the PeV energy range and above. 
The present formulation of this idea is often referred to as {\it the SuperNova Remnant (SNR) paradigm for the origin of CRs}, because galactic CRs are believed to be accelerated via first order Fermi mechanism operating at SNR shocks \citep[e.g.][]{hillas,eveline}. 

The success of this paradigm relies on several facts. First of all, SNRs can provide the power needed to sustain the CR flux at the observed level, if some fraction (about $10 ...  30\%$) of their kinetic energy is somewhat converted into relativistic particles \citep[e.g.][and references therein]{seo}. Second, diffusive shock acceleration can operate at SNR shocks \citep[for a review see][]{drury83}, and this provides a viable mechanism to accelerate CRs. Diffusive shock acceleration is expected to accelerate particles with power law spectra whose slope, once the effects of CR propagation in the Galaxy are taken into account, roughly matches the one of the CR spectrum as it is observed at the Earth \citep[e.g.][]{seo,bellsteep,damiano}. Finally, during the acceleration process CRs can amplify the magnetic field at shocks via various plasma instabilities \citep[e.g.][]{bell78,bell04,luke}. Observational evidence for such an amplification has been found \citep[see][for a review]{jaccoreview}, and it is likely that the magnetic field strength at SNR shocks may grow up to a level that allows the acceleration of particles up to PeV energies and more. 
All these things support the idea that SNRs indeed are the sources of CRs, but it has to be kept in mind that an unambiguous and conclusive proof of such a statement is still missing.

The acceleration of CRs must be accompanied by the production of gamma--rays. This radiation is the result of the decay of neutral pions generated in hadronic interactions between the CRs and the ambient gas \citep{steckerbook,felixbook}. It was shown by \citet{dav} and \citet{nt} that if SNRs indeed are the sources of CRs, then their gamma--ray emission must be strong enough to be detected by Cherenkov telescopes of current generation. The detection of several SNRs in TeV gamma rays nicely fits with these earlier predictions, but it cannot be considered a proof of the fact that SNRs can accelerate CRs. This is because electrons can as well be accelerated at shocks, and their inverse Compton emission can also account for the observed TeV radiation \citep[e.g.][]{felixreview,mereview,jimreview}. The ambiguity between the hadronic or leptonic origin of the gamma--ray emission observed from SNRs is the main obstacle in proving (or falsifying) the SNR paradigm for the origin of CRs. 

Multi--wavelength observations of SNRs, from the radio band to the very high energy gamma--ray domain, can help in solving such degeneracy. This is generally done on a case by case basis, i.e. multi--wavelength data are collected for a specific SNR, and hadronic and/or leptonic models are fitted to data \citep[see e.g.][and references therein]{don,heinz,vladimir,giovanni}. In some cases, it has been possible to ascribe quite confidently the gamma--ray emission either to a hadronic or a leptonic mechanism \citep{fermitycho,giovannitycho,fermiRXJ,donRXJ}, while for other cases the situation still remains ambiguous.

In this paper, we follow a different approach and, instead of considering one specific object, we investigate the gamma--ray properties of SNRs as a class of objects. We start from the assumption that SNRs are the sources of CRs. This assumption, together with the knowledge of the supernova rate in the Galaxy, allows us to infer the typical CR acceleration efficiency per SNR. Then, by means of a Monte Carlo method we simulate the location and time of explosion of all the supernovae in the Galaxy. Finally, from the information on the gas density, taken from galactic surveys of CO and HI lines (that trace molecular and atomic Hydrogen, respectively), it is possible to estimate the hadronic gamma--ray emission from each simulated SNR.
A leptonic contribution is added, by treating the electron--to--proton ratio $K_{ep}$ as a free parameter of the model.
Following this procedure it is possible to build mock--catalogues of TeV--bright SNRs that can then be compared with the data coming, for example, from the H.E.S.S. survey of the galactic plane \citep{scan1}. Our results show that expectations match quite well current observations, providing an additional and novel consistency check for the SNR paradigm for the origin of galactic cosmic rays.

The paper is structured as follows: in Section~\ref{sec:model} we describe a procedure to estimate the gamma--ray emission from an individual SNR at a given stage of evolution, while in Section~\ref{sec:montecarlo} a Monte Carlo approach is adopted to simulate the position and time of explosion of all the supernovae that exploded in the Galaxy. These results are then used in Section~\ref{sec:montecarlo} to estimate the number of SNR detectable in the Galaxy at a given gamma--ray flux. In Section~\ref{sec:hess} a comparison with existing data (mainly from the survey of the galactic plane performed by the H.E.S.S. collaboration) is performed. Finally, we discuss the results and conclude in Sections~\ref{sec:discussion}~and~\ref{sec:conclusions}.

\section{A model for cosmic ray acceleration and gamma ray production in supernova remnants}
\label{sec:model}

In this section we develop a model that couples the dynamical evolution of a SNR with the particle acceleration operating at the shock.
The aim of the model is to obtain predictions for the gamma--ray emission from individual SNRs.

\subsection{Dynamical evolution of supernova remnants} 
\label{sec:evolution}

In order to determine the time evolution of the SNR shock radius and velocity, we follow the approach outlined in \citet{pz03,pz05}, where a significant contribution of CRs to the pressure behind the SNR shock has been assumed. 

Let us consider first the case of a thermonuclear, type Ia, supernova. The time evolution of the shock radius $R_{sh}$ and velocity $u_{sh}$ in the ejecta--dominated phase are described by the following self--similar expressions \citep{chevalier,pz05}:
$$
R_{sh} = 5.3 \left( \frac{{\cal E}_{51}^2}{n_0 ~ M_{ej,\odot}} \right)^{1/7} t_{\rm kyr}^{4/7} ~~ {\rm pc}
$$
\begin{equation}
\label{eq:ejIa}
u_{sh} = 3.0 \times 10^3 \left( \frac{{\cal E}_{51}^2}{n_0 ~ M_{ej,\odot}} \right)^{1/7} t_{\rm kyr}^{-3/7} ~~ {\rm km/s}
\end{equation}
where ${\cal E}_{51}$ is the supernova explosion energy in units of $10^{51}$~erg, $n_0$ is the ambient gas number density in cm$^{-3}$, $M_{ej,\odot}$ is the mass ejected in the explosion in solar mass units, and $t_{\rm kyr}$ is the time after explosion expressed in kilo--years.
To describe the SNR evolution during the adiabatic phase it is convenient to use the expressions \citep{truelove,pz05}:
\begin{equation}
R_{sh} = 4.3 \left( \frac{{\cal E}_{51}}{n_0} \right)^{1/5} t_{\rm kyr}^{2/5} 
\left( 1 - \frac{0.06 ~ M_{ej,\odot}^{5/6}}{{\cal E}_{51}^{1/2} n_0^{1/3} t_{\rm kyr}} \right)^{2/5} {\rm pc} 
\end{equation}
$$
u_{sh} = 1.7 \times 10^3 \left( \frac{{\cal E}_{51}}{n_0} \right)^{\frac{1}{5}} t_{\rm kyr}^{-3/5}
\left( 1 - \frac{0.06 ~ M_{ej,\odot}^{5/6}}{{\cal E}_{51}^{1/2} n_0^{1/3} t_{\rm kyr}} \right)^{-\frac{3}{5}} {\rm km/s} 
$$
which connect smoothly with Equations~\ref{eq:ejIa} at a time $t_0 \sim 260 (M_{ej,\odot}/1.4)^{5/6} {\cal E}_{51}^{-1/2} n_0^{-1/3}$~yr, and tend to the exact Sedov--Taylor solution \citep{sedov,taylor} for $t \gg t_0$.
We follow the SNR evolution until the time $t_{rad} \approx 3.6 \times 10^4 {\cal E}_{51}^{3/14} n_0^{-4/7}$~yr, which marks the transition to the radiative phase \cite{cioffi}.

Different expressions need to be adopted to describe the evolution of a core--collapse supernova, whose shock propagates in the wind blown bubble generated by the wind of the progenitor star. Following \citet{pz05} we divide the wind blown bubble into two regions: a dense red--supergiant wind and a tenuous hot bubble which has been inflated by the wind of the massive progenitor star in main sequence. The red--supergiant wind is assumed to be spherically symmetric with velocity $u_w = 10^6 u_{w,6}$~cm/s, mass loss rate $\dot{M} = 10^{-5} \dot{M}_{-5} M_{\odot}$/yr, and density $n_w = \dot{M}/4 \pi m_a u_w r^2$, where $m_p$ is the proton mass and $m_a \approx \mu m_p$ is the mean interstellar atom mass (here we adopt $\mu \approx 1.4$) and $r$ is the distance from the star. The radius of the wind is fixed to $R_w \approx 2$~pc, since its exact location does not affect significantly the results. The radius of the hot bubble is $R_b = 28 (L_{36}/\mu n_0)^{1/5} t_{\rm Myr}^{3/5}$~pc, where $L_{36}$ is the main--sequence star wind power in units of $10^{36}$~erg/s, and $t_{\rm Myr}$ is the wind lifetime in units of mega--years. The density inside the bubble is $n_b = 0.01 (L_{36}^{6} n_0^{19} t_{\rm Myr}^{-22})^{1/35}$~cm$^{-3}$ \citep{weaver}. Here we assume $t_{\rm Myr}$ to be of the order of several Myr, which corresponds to the duration of the main sequence phase of very massive stars \citep{longair}. 

In such a structured interstellar medium, the evolution of the SNR shock during the ejecta dominated phase is described by \citep{chevalier,pz05}:
$$
R_{sh} = 7.7 \left( \frac{{\cal E}_{51}^{7/2} u_{w,6}}{\dot{M}_{-5} M_{ej,\odot}^{5/2}} \right)^{1/8} t_{\rm kyr}^{7/8} ~~ {\rm pc}
$$
\begin{equation}
\label{eq:ejII}
u_{sh} = 6.6 \times 10^3 \left( \frac{{\cal E}_{51}^{7/2} u_{w,6}}{\dot{M}_{-5} M_{ej,\odot}^{5/2}} \right)^{1/8} t_{\rm kyr}^{-1/8} ~~ {\rm km/s}
\end{equation}
Fairly accurate expressions that describe the evolution of a SNR in the adiabatic phase can be obtained, in this case, by adopting the thin--shell approximation \citep[e.g.][]{ostriker,bisnovati}, i.e. the assumption that the gas swept up by the SNR shock is concentrated in a thin layer behind the shock front. The following equations can be derived, where the shock speed and SNR age are parametrized as functions of the shock radius \citep{pz05}:
\begin{equation}
\begin{split}
u_{sh}(R_{sh}) = \frac{\gamma_{ad}+1}{2} \Biggl[ \frac{12 (\gamma_{ad}-1) {\cal E}}{(\gamma_{ad}-1) M^2(R_{sh}) R_{sh}^{6 (\gamma_{ad}-1)/(\gamma_{ad}+1)}} \\
\times \int_0^{R_{sh}} {\rm d}r r^{6 (\gamma_{ad}-1)/(\gamma_{ad}+1) - 1} M(r) \Biggr]^{1/2}
\end{split}
\nonumber
\end{equation} 
\begin{equation}
\label{eq:adII}
t(R_{sh}) = \int_0^{R_{sh}} \frac{{\rm d}r}{u_{sh}(r)}
\end{equation}
In the expressions above, $\gamma_{ad}$ is the gas adiabatic index, and $M(R_{sh})$ is the total gas mass inside the SNR shock (ejecta+swept--up). Equations~\ref{eq:ejII} and \ref{eq:adII} are fitted together at the transition between the ejecta--dominated and adiabatic phases. 
We follow the evolution of the SNR until the Mach number drops to $\approx$~3, or, if shorter, until the time at which the SNR shock impacts onto the unperturbed (and much denser than the gas inside the bubble) interstellar medium and becomes quickly radiative.

The internal structure of the SNR is determined by adopting the linear velocity approximation \citep{ostriker}, in which the gas velocity profile for $r < R_{sh}$ is given by:
\begin{equation}
\label{eq:u}
u(r,t) = \left( 1-\frac{1}{\sigma} \right) u_{sh}(t)  \frac{r}{R_{sh}(t)}
\end{equation}
where $\sigma$ is the shock compression ratio. In the absence of CR acceleration, for a strong shock $\sigma = 4$. Conversely, if the CR pressure at the shock cannot be neglected, the compression ratio increases due to the reaction of CRs onto the shock structure. To date, there is a general consensus on the fact that the increase of the shock compression in SNRs is quite moderate, with $\sigma \lesssim 10$, where the larger values are obtained for extremely high acceleration efficiencies only \citep[e.g.][]{damianocompression,donRXJ,heinz,vladimir}. Here we follow \citet{vladimir2012} and adopt $\sigma = 6$ as a reference value. Equation~\ref{eq:u} can be combined with the gas continuity equation to determine the density profile inside the SNR \citep{ostriker}. Such profile will be used in the following to compute the hadronic contribution to the gamma--ray emission of SNRs.

\subsection{Cosmic ray acceleration at supernova remnant shocks and related gamma--ray emission}
\label{sec:crgamma}

Let $f(r,p,t)$ be the CR particle distribution function at a given time $t$ after the supernova explosion and at a given position $r < R_{sh}$ inside the SNR. Here, $p$ is the particle momentum. We assume that particles are accelerated at the shock with a power law spectrum so that $f_0(p,t) \equiv f(R_{sh},p,t) = A(t) p^{-\alpha}$, where $\alpha$ is treated as a free parameter with values in the range $\alpha \gtrsim 4$ \citep[see e.g.][]{vladimirdrift,damiano}. Values significantly larger than $\alpha = 4$ are needed to reproduce the slope of the CR spectrum observed at the Earth. In the following we will consider two representative cases: $\alpha = 4.4$ (soft spectrum) and $\alpha = 4.1$ (hard spectrum). The normalization $A$ of the CR spectrum is computed by assuming that the CR pressure at the shock is equal to some fraction $\xi_{CR}$ of the shock ram pressure:
\begin{equation}
\label{eq:xi}
P_{CR}^0 = \xi_{CR} ~  \varrho_{up} ~ u_{sh}^2
\end{equation}
with $\varrho_{up}$ representing the gas mass density upstream of the shock. The parameter $\xi_{CR}$ is a way to express the acceleration efficiency at the shock. For spectra steeper than $\alpha = 4$ the dominant contribution to the CR pressure comes from particles with momentum $p \approx m_p c$. On the other hand, the maximum momentum $p_{max}$ of the accelerated particles is determined here by the 
equality:
\begin{equation}
\label{eq:confinement}
l_d = \frac{D(p_{max})}{u_{sh}} \approx \zeta R_{sh}
\end{equation}
where $D$ represents the momentum dependent diffusion coefficient for CRs upstream of the shock.
In this way, $p_{max}$ is defined as the momentum for which the diffusion length of particles ahead of the shock $l_d$ equates some fraction $\zeta$ of the shock radius. Particles with larger momentum are characterized by larger diffusion length and are assumed to escape the SNR. Studies of particle acceleration and escape from shocks suggest that $\zeta \approx 0.05...0.1$ \citep[e.g.][and references therein]{zirakashviliptuskin08}. In the following we will adopt the value $\zeta = 0.1$.

The particle diffusion coefficient at the shock depends on the magnetic field strength and on its turbulent structure. If the CR acceleration efficiency is high, the intensity of the turbulent magnetic field at shocks is expected to be strongly amplified with respect to the typical value found in the interstellar medium, and the diffusion coefficient is suppressed accordingly. The field amplification might be due to a CR current driven instability \citep{bell04} or to resonant streaming instability \citep{lagage}. The determination of the CR diffusion coefficient in the turbulent magnetic field at shocks is still an open issue. Here we make the assumption that, in the presence of efficient field amplification, the CR diffusion coefficient is of the Bohm type in the amplified field. The expression for the Bohm diffusion coefficient is $D = R_L c/3$ and $R_L = pc/qB$ is the particle Larmor radius. In the expressions above, $c$ is the speed of light, $q$ the elementary charge, and $B$ the amplified magnetic field. To estimate the value of the amplified field we rely on the interpretation of X--ray data from SNRs made by \citet{heinzB}. Starting from the observations of X--ray filaments in several young SNRs, they estimated the magnetic field intensity just downstream of the shock wave. They found that a fraction $\xi_B \approx 3.5\%$ of the shock ram pressure $\varrho_{up} u_{sh}^2$ is, on average, converted into magnetic field. By assuming that this fraction remains constant during the SNR lifetime, an expression for the amplified magnetic field strength immediately downstream of the shock can be derived, and reads:
\begin{equation}
\label{eq:amplifiedB}
B_{down} = B_0 ~ \sigma ~ \sqrt{ \left( \frac{u_{sh}}{v_d} \right)^2 + 1}
\end{equation}
where $B_0 \approx 5~ \mu$G is the magnetic field in the interstellar medium and $v_d$ is a velocity that defines the importance of wave damping in limiting the field amplification \citep[e.g.][]{damianodamping}. In Eq.~\ref{eq:amplifiedB}, the term under the square root represents the amplification due to the CR instability operating upstream of the shock, while the factor $\sigma$ mimics the effect of the field compression at the shock. For shock velocities larger than $v_d$ the damping of the magnetic turbulence at the shock is negligible, and thus the field can be effectively amplified. On the other hand, damping dominates for smaller velocities.
Following \citet{vladimir} we adopt for $v_d$ the following expression:
\begin{equation}
v_d = \left( \frac{\sigma^2 B_0^2}{8 \pi \xi_B \varrho_{up}} \right)^{1/2} \approx 0.2 \times 10^8  ~ n_0^{-1/2}  {\rm cm/s}
\end{equation}
which is in substantial agreement with the results of more sophisticated studies \citep[e.g.][]{pz03}.
Moreover, we adopt here a diffusion coefficient for CRs at the shock of the form \citep{vladimir2012}:
\begin{equation}
\label{eq:diffusioncoefficient}
D = D_B \left( 1+ \frac{v_d^2}{u_{sh}^2}\right)^g
\end{equation}
where $D_B$ is the Bohm diffusion coefficient and the parameter $g$ depends on the nature of the dominant damping mechanism. 
We fix here $g = 3$, as appropriate for a Kolmogorov type of non--linear damping \citep{pz03}. Equation~\ref{eq:diffusioncoefficient} is valid in both regimes $u_{sh} > v_d$ and $u_{sh} < v_d$: in the former, the diffusion coefficient coincides with the Bohm one, while in the latter it is significantly larger than that due to wave damping.  

It is evident from Eq.~\ref{eq:amplifiedB} that the amplified field, being proportional to the shock speed, decreases with time, as long as $u_{sh} \gg v_d$:
\begin{equation}
\label{eq:Bdown}
B_{down} \approx ~ 140 ~ \left( \frac{u_{sh}}{1000~{\rm km/s}} \right) ~ n_0^{1/2} ~ \mu {\rm G}
\end{equation} 
After that, field amplification becomes inefficient and the magnetic field downstream of the shock stays constant in time and is equal to $\sigma B_0$ (i.e. only the compression of the field at the shock is taken into account). This fact, once combined with Eq.~\ref{eq:confinement}, implies that the maximum momentum of the protons that can be confined and accelerated at the shock decreases with time. As an illustrative example, for a SNR expanding adiabatically in an uniform medium Equations~\ref{eq:amplifiedB} and \ref{eq:Bdown} give, for $u_{sh} \gg v_d$, $B \propto u_{sh} \propto t^{-3/5}$, which corresponds to (Eq.~\ref{eq:confinement})
$p_{max} \propto t^{-4/5}$. A quantitative expression for the maximum energy $E_{max} = p_{max} c$ can easily be computed and reads:
\begin{equation}
\label{eq:Emax1}
E_{max} \approx ~ 280 ~ {\cal E}_{51}^{3/5} n_0^{-1/10} t_{kyr}^{-4/5} ~ {\rm TeV}
\end{equation}
which implies that at an age of few hundred years SNRs are capable to accelerate particles up to the PeV domain. 
In the opposite situation in which the magnetic field at the shock is not amplified and stays constant in time, the maximum momentum of protons exhibits a very slow decline in time. This happens at late times and can be described as:
$$
E_{max} \approx ~17~ \left( \frac{{\cal E}_{51}}{n_0} \right)^{2/5} \left( \frac{t_{kyr}}{30} \right)^{-1/5} \times ~~~~~~~~~~~~~~
$$
\begin{equation}
\label{eq:Emax2}
\times \left[ 1+ 0.1 \left( \frac{t_{kyr}}{30~n_0~{\cal E}_{51}} \right)^{3/5} \right]^{-3} {\rm TeV}
\end{equation}
Similar estimates can be obtained also for supernovae exploding in a structured medium (i.e. wind plus bubble), though analytic expressions cannot be written in this case.


The spatial distribution of CRs inside the remnant can be computed by solving the transport equation:
\begin{equation}
\label{eq:transport}
\frac{\partial f}{\partial t} + u \nabla f - \nabla D \nabla f - \frac{p}{3} \nabla u \frac{\partial f}{\partial p} = 0
\end{equation}
where $D$ is a momentum dependent diffusion coefficient for CRs. Particles with momenta smaller than $p_{max}$ are expected to be well confined within the SNR. Thus, Equation~\ref{eq:transport} can be solved by dropping the diffusion term $\nabla D \nabla f$, which is expected to be negligible when compared to the advection term $u \nabla f$. The solution of this differential equation can be found by using the method of the characteristics and adopting the boundary condition $f(R_{sh},p,t) = f_0(p,t)$.

The acceleration of electrons at shocks proceeds at the same rate as for protons, but their spectrum is different because electrons suffer synchrotron and inverse Compton losses, while protons are virtually loss--free. At low energies, where energy losses can be neglected, the acceleration of electrons and protons at the shock proceeds in an identical manner, implying that the same spectral shape is expected for both species. Thus, it is possible to introduce a parameter $K_{ep}$, generally believed to be much smaller than unity, that describes the ratio between the electron and proton spectra at low energies.

The maximum energy of the electrons accelerated at a shock can be obtained by equating the acceleration rate at the shock to the synchrotron energy loss time. To compute this, we follow the approach described in \citet{giulia}, which gives:
\begin{equation}
\label{eq:Emaxe}
E_{max}^e \approx ~7.3 ~\left( \frac{u_{sh}}{1000~{\rm km/s}} \right) \left( \frac{B_{down}}{100 ~ \mu {\rm G}} \right)^{-1/2} {\rm TeV}
\end{equation}
At this energy, a cutoff appears in the electron spectrum, with shape $\propto \exp{[-(E/E_{max}^e)^2]}$ \citep{volodiaelectrons}.

Electrons are accelerated very quickly, over time scales significantly shorter than the synchrotron energy loss time which, for 10 TeV electrons in a 100 $\mu$G field is of the order of a century. After being accelerated they are advected downstream of the shock, where they continue to lose enegy mainly through synchrotron radiation, with a characteristic time: 
\begin{equation}
\tau_{syn} \approx ~ 1.8 \times 10^3 \left( \frac{E_e}{{\rm TeV}} \right)^{-1} \left( \frac{B_{down}}{100~\mu{\rm G}} \right)^{-2} {\rm yr}
\label{eq:tausyn}
\end{equation}
where $E_e$ is the electron energy. The energy loss time decreases with particle energy and this implies that an energy $E_{break}^e$ exists above which the loss time is shorter than the SNR age $\tau_{age}$. Above such energy, the electron spectrum is shaped by radiative losses and steepens by one power in momentum \citep[see e.g.][]{giovannitycho}. In fact, since the SNR is expanding also adiabatic losses have to be taken into account, with a rate $\tau_{ad} = R_{sh}/u_{sh}$. After a comparison with the work of \citet{finke} we found that an appropriate expression for $E_{break}^e$ can be found by solving the equation $\tau_{age}^{-1} = \tau_{syn}^{-1} + \tau_{ad}^{-1}$.


Two things have to be noted. First, if the magnetic field is not strong enough, $E_{break}^e$ can easily become larger than $E_{max}^e$. In this case, no break appears and the electron spectrum has the same shape as the one of protons up to $E_{max}^e$. Secondly, in some situations Eq.~\ref{eq:confinement} can be more stringent than Eq.~\ref{eq:Emaxe} (i.e. the acceleration of electrons is limited by escape rather that by energy losses), and in this case the former is used to estimate the energy of the cutoff in the electron spectrum.
 
The last missing piece of information is the value of the parameter $\xi_{CR}$, defined in Eq.~\ref{eq:xi}, which represents the particle acceleration efficiency at the shock. In order to estimate this parameter, we assume that SNRs are the sources of galactic CRs. The estimated CR luminosity of the Galaxy is of the order of $L_{CR}^{MW} \approx 10^{41}$~erg/s and this number is quite stable with respect to the assumptions made to derive it \citep{dogiel,CRMW}\footnote{\citet{CRMW} give luminosities in the interval $6.6...8.1 \times 10^{40}$~erg for CRs in the energy range 0.1...100 GeV. \citet{dogiel} give values of $L_{CR}^{MW}$ in the range $0.5...1 \times 10^{41}$~erg.}. By combining this information with the estimated rate of supernovae in the Galaxy $\nu_{SN} \approx 3$/century \citep[e.g.][and references therein]{SNrate} it is possible to obtain the average fraction of the supernova explosion energy that needs to be converted into CRs in order to provide the required power $L_{CR}^{MW}$. This gives:
\begin{equation}
\label{eq:efficiency}
\eta_{CR} = \frac{L_{CR}^{MW} \nu_{SN}}{{\cal E}_{SN}}  \approx 0.1 \left( \frac{L_{CR}^{MW}}{10^{41} {\rm erg}} \right) \left( \frac{\nu_{SN}}{0.03 {\rm yr}^{-1}} \right) \left( \frac{{\cal E}_{SN}}{10^{51} {\rm erg}} \right)^{-1}
\end{equation}
The parameter $\eta_{CR}$ represents the global (i.e. integrated over the whole SNR lifetime) CR output from a single SNR, while the parameter $\xi_{CR}$ that appears in Eq.~\ref{eq:xi} measures the instantaneous (i.e. at a specific time) acceleration efficiency at the shock. Inspired by the results presented by \citet{vladimir2012} and \citet{damiano} we will assume in the following that $\xi_{CR}$ remains constant over the SNR lifetime up to the end of the Sedov phase and that $\xi_{CR} \approx \eta_{CR} \approx 0.1$.
However, the value of the CR acceleration efficiency which is expected from theoretical studies of non--linear shock acceleration is generally significantly larger that the modest $\eta_{CR} \approx 10$\% needed to sustain the observed flux of galactic CRs.
One way to solve this apparent discrepancy is to assume that CRs can be accelerated with high efficiency (and thus modify the shock structure) only in a small fraction of the shock surface \citep[see e.g.][]{heinz1006,BKV09,vladimir2012}.
Thus, here the value $\xi_{CR} \approx 0.1$ has to be considered as an average over the whole SNR shock surface. 

The procedure described in this section allows to determine the spectrum and the spatial distribution of CRs inside a SNR.
By combining these results with the spatial distribution of the gas inside the SNR as obtained in Sec.~\ref{sec:evolution}, the gamma--ray luminosity from a given SNR can be computed, by adding the hadronic contribution from proton--proton interactions \citep{kelner} to the leptonic one from inverse Compton scattering off photons in the cosmic microwave background \citep{gould}. While computing the gamma--ray emission from proton--proton interactions the results from \citet{kelner} have been multiplied by a factor of $\approx 1.8$ to take into account the presence of nuclei heavier than Hydrogen in both ambient gas and CRs \citep{mori}.

\section{On the number of supernova remnants detectable in the TeV energy domain: a Monte Carlo approach}
\label{sec:montecarlo}

In this section we describe the simulation procedure adopted to predict the number of galactic SNRs with a given gamma--ray flux. Since this work is focused on the TeV energy domain in the following we will always consider integral fluxes computed above a photon energy of 1 TeV.

A Monte Carlo approach is used to simulate the time of explosion of all the supernovae in the Galaxy (i.e. the age of all the SNRs in the Galaxy). This has been done by assuming a supernova explosion rate constant in time and equal to $\nu_{SN} = 3$/century \citep{SNrate}. This implies an acceleration efficiency at the shock of the order of $\eta_{CR} \approx 10\%$ (see Eq.~\ref{eq:efficiency}). Different values of $\nu_{SN}$ will be also discussed in the following. Once the age of a SNR is known, its location within the Galaxy is determined by following the prescription described in \citet{faucher}. This consists in assuming that the radial distribution of SNRs in the Galaxy follows the distribution of pulsars, as determined by \citep{PSR, lorimer}. In addition to that, four spiral arms are considered, each arm following a logarithmic spiral shape (see Table~2 in \citealt{faucher}). The width of the arms has been modeled as in \citet{blasiamato}. To determine the height above (or below) the galactic plane of a SNR, we assume that the vertical distribution of supernovae follows the one of the gas. We use the vertical distribution of molecular and atomic Hydrogen for core--collapse and type Ia supernovae, respectively \citep{shibata}, which implies that the distribution of type Ia supernovae has a height above the disk which is significantly larger than the one of core--collapse supernovae. In the absence of a detailed knowledge of the spatial distribution of supernovae of a given type, this assumption mimics the fact that core--collapse supernovae are expected to explode in dense star forming regions, while thermonuclear ones can be found also in low density regions.

\begin{table}
\centering
\begin{tabular}{c c c c c c}
\hline 
Type & ${\cal E}_{51}$ & $M_{ej,\odot}$ & $\dot{M}_{-5}$ & $u_{w,6}$ & Rel. rate\\
\hline 
\hline
Ia & 1 & 1.4  & -- & -- & 0.32\\
IIP & 1 & 8 & 1 & 1  & 0.44 \\
Ib/c & 1 & 2 & 1 & 1 & 0.22 \\
IIb & 3 & 1 &  10 & 1 & 0.02 \\
\hline
\end{tabular}
\caption{Supernova parameters adopted in the simulation: supernova type (column 1), explosion energy in units of $10^{51}$~erg (column 2), mass of ejecta in solar masses (column 3), the wind mass loss rate in $M_{\odot}$/yr (column 4), the wind speed in units of 10~km/s (column 5), and the relative explosion rate (column 6). Values from \citet{seo}.}
\label{table1}
\end{table}

The dynamical evolution of each simulated SNR is then determined as explained in Sec.~\ref{sec:evolution}. The evolution depends mainly on the value of the ambient density at the location of the SNR and on the supernova type. To determine the value of the ambient density we use the three--dimensional distributions (galactic latitude, longitude, and radial velocity) of atomic ($H$I) and molecular ($H_2$) Hydrogen given by \citet{HI, H2}. The three--dimensional spatial distribution of the gas (i.e. the conversion from radial velocity to distance) is computed as in \citet{sabrina}. Four types of supernovae are considered: Ia, IIP, Ib/c, and IIb, with relative rates 0.32, 0.44, 0.22, and 0.02, respectively \citep{seo}. The parameters used for each supernova type to compute the SNR dynamical evolution are listed in Table~\ref{table1}. The gamma--ray emission from each SNR is then computed as described in Sec.~\ref{sec:crgamma}.

The procedure described in this Section can be used to simulate the number of SNRs that one can expect to detect with a Cherenkov telescope with a given sensitivity. This will be done in the next Section, where the prediction from our Monte Carlo will be compared with the data from the survey of the galactic plane performed by the H.E.S.S. collaboration. Before doing that, we compute here the total number of SNRs in the Galaxy which are expected to emit gamma-rays above a given flux. All the results reported in the following have been obtained by averaging 1000 Monte Carlo realizations of the Galaxy.
 
\begin{figure}
\includegraphics[width=.5\textwidth]{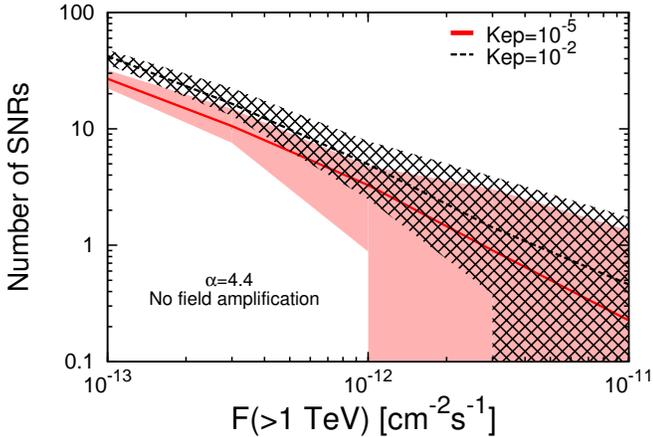}
\caption{Number of SNRs in the Galaxy with integral gamma--ray flux above $F(> 1~{\rm TeV})$. The spectral slope of accelerated particles at the shock is $\alpha = 4.4$, the electron--to--proton ratio is $K_{ep} = 10^{-5}$ and $10^{-2}$ for the red and black curve, respectively (models M1 and M2 in Table~\ref{tab:models}).}
\label{fig:nonamplified}
\end{figure}

We first consider a situation in which the magnetic field at the shock is not amplified. In this case, particles are accelerated at a shock characterized by an upstream magnetic field of $B_{up} = B_0  \approx 5 ~ \mu {\rm G}$ and a downstream magnetic field of $B_{down} = \sigma B_0 \approx 30 ~\mu {\rm G}$. In Figure~\ref{fig:nonamplified} we plot the number of SNRs in the Galaxy which are expected to have an integral gamma--ray flux above a given value. Integral fluxes above 1~TeV are considered. The red solid line shows our prediction for a soft spectrum of accelerated CRs with slope $\alpha = 4.4$. A very small electron--to--proton ratio $K_{ep} = 10^{-5}$ is assumed, and this insures that for all the SNRs the hadronic emission largely dominates over the leptonic one. The shaded red region around the curve shows the fluctuations of the results due to the stochasticity of the process. To estimate that, histograms representing the number of realizations that correspond to a given number of detections have been produced and fitted with continuous functions. The shaded region represents the interval within which 68.2\% of the area below the fitting function is contained.
The black dashed curve (as well as the shaded black region) has been computed, instead, by assuming a high electron--to--proton ratio $K_{ep} = 10^{-2}$. The number of SNRs expected at each flux is increased by a factor of $\gtrsim 1.5$ with respect to the case $K_{ep} = 10^{-5}$, and this is due to the fact that the leptonic contribution is no longer negligible. The fact that the increase in the number of TeV--bright SNRs is modest but not negligible (it is indeed close to a factor of 2) indicates that for $K_{ep} = 10^{-2}$ the number of SNRs for which the hadronic emission dominates the gamma--ray flux is of the same order of the number of SNRs for which the leptonic emission dominates.

The number of SNRs with an integral flux greater than 1\% of the Crab ($\approx 2.3 \times 10^{-13}$cm$^{-2}$s$^{-1}$, \citealt{hesscrab} ), a representative flux sensitivity for deep pointed observations with Cherenkov telescopes of current generation\footnote{This is true as long as point sources are considered. For extended sources, as SNR often are, the sensitivity is worse, and roughly scales as the source size. So, the numbers reported here have to be considered only as reference (and quite optimistic!) values. A full discussion of this issue can be found in Section~\ref{sec:hess} below, where a comparison with existing data is attempted.}, is $\approx 13$ and $\approx 21$ for the red and black curve, respectively. 
On the other hand, the probability to detect very bright SNRs with fluxes of the order of $\approx 10^{-11}$~cm$^{-2}$~s$^{-1}$ is small, but not completely negligible. For this value of the integral flux, the mean values for the expected number of gamma--ray SNRs are $\approx 0.2$ and $\approx 0.5$ for the red and black curve, respectively, while the shaded regions (representing one standard deviation) extends up to $\approx 1.3$ and $\approx 1.7$. This is still consistent, though in some tension with the fact that two SNRs have been detected at such flux levels: RX J1713.7--3946, with an integral flux equal to $F(>1~{\rm TeV}) \approx 1.6 \times 10^{-11}$~cm$^{-2}$s$^{-1}$ \citep{hessRXJ} and Vela Jr, with an integral flux equal to $F(>1~{\rm TeV}) \approx 1.5 \times 10^{-11}$~cm$^{-2}$s$^{-1}$ \citep{hessVelaJr}. Unfortunately, a more quantitative comparison between our predictions and available data is, at this stage, not easy to be performed, due to the lack of a complete catalogue of TeV gamma--ray bright SNRs. This point will be extensively discussed in Section~\ref{sec:hess}.
 
\begin{figure}
\includegraphics[width=.5\textwidth]{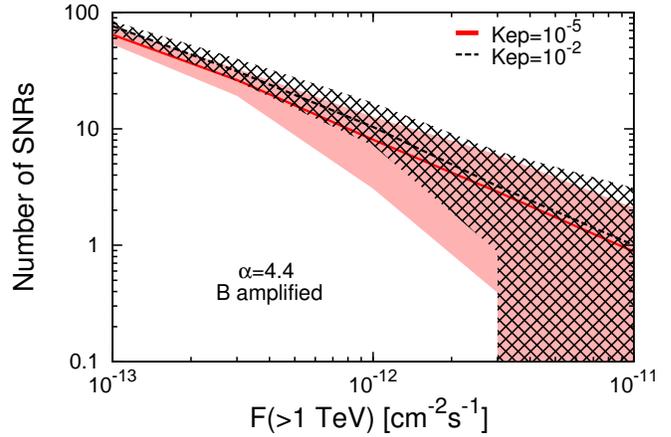}
\caption{Same as in Fig.~\ref{fig:nonamplified} with the only exception that an amplified field has been considered (models M3 and M4 in Table~\ref{tab:models}).}
\label{fig:amplified}
\end{figure}

A much more plausible scenario, supported by both theory \citep{bell04} and observations \citep{jaccoreview}, is the one in which the magnetic field at the shock is substantially amplified due to CR-induced instabilities that may operate in the shock precursor. In this case the values of the magnetic field and of the maximum energy of accelerated protons can reach values up to hundreds of microGauss and PeV energies or even more, respectively. These high values are achieved early in the evolution of the SNR and then gradually decrease with time as the shock slows down. A plausible parametrization of this behavior has been described in Section~\ref{sec:crgamma}.
The expected number of SNRs with integral flux above $F(>1~{\rm TeV})$ is  shown in Figure~\ref{fig:amplified} for $K_{ep} = 10^{-5}$ and $10^{-2}$ (red and black curve, respectively). The dashed regions have the same meaning as in Figure~\ref{fig:nonamplified}. 
The number of SNRs with flux above the 1\% of the Crab is $\approx 32$ and $\approx 39$ for the red and black curve, respectively. The mean value for the number of very bright SNRs, with fluxes above $10^{-11}$cm$^{-2}$s$^{-1}$ is, for both curves, $\approx 1$, in closer agreement with the detection of the two very bright SNRs.

The first thing to be noted is that for $K_{ep} = 10^{-5}$ (i.e. no leptonic contribution to the gamma--ray emission) the number of SNRs at a given flux is larger (by roughly a factor of $\approx 2...4$) when the magnetic field is amplified. This can be seen by comparing the red lines in Figures~\ref{fig:nonamplified}~and~\ref{fig:amplified}. The reason for this is the fact that, in order to detect the hadronic interaction of a SNR above a photon energy of 1~TeV, the underlying proton spectrum must extend up to energies significantly larger than $\approx 10$~TeV, because these are the particles that produce the photons with energy in excess of 1 TeV. The maximum energy of the protons accelerated at the shock is larger if the field is amplified and thus, in this case, SNRs remain visible above 1 TeV for a longer time. This explains why one expects to see more gamma--ray SNRs if the magnetic field is amplified (even if the acceleration efficiency is the same in the two cases).

Another thing to be noted is that, if the field is amplified, there is not much difference in our predictions if a low or a high value of the electron--to--proton ratio $K_{ep}$ is adopted. In fact, the black and red curves in Figure~\ref{fig:amplified}, which refer to $K_{ep} = 10^{-2}$ and $10^{-5}$, are virtually identical for gamma--ray fluxes larger than $\gtrsim 10^{-12}$cm$^{-2}$s$^{-1}$, and remain comparable for all the values of the gamma--ray fluxes (the difference between the two curves is always less than $\approx 30 \%$). This implies that the leptonic gamma--ray emission from SNRs never plays a crucial role. If the field is amplified, electrons suffer severe synchrotron energy losses, and as a consequence of that, their spectrum exhibit a break at an energy which can be computed by equating the energy loss time with the SNR age (see the discussion following Eq.~\ref{eq:tausyn}). For large values of the field (few hundreds microGauss) and typical SNR ages of thousands of years the break appears at TeV energies. The electron spectrum below the break is identical to the proton one (i.e. it is a power law in momentum with slope 4.4) while above the break the spectrum steepens by one power in momentum. Such steepening suppresses the leptonic emission in the TeV domain, and explains why the parameter $K_{ep}$ virtually plays no role in this case.
 
In computing the curves in Figure~\ref{fig:amplified} we assumed that the magnetic field downstream of the shock is the amplified one, as determined by Eq.~\ref{eq:Bdown}. This might not be the case if the turbulent magnetic field is damped downstream of the shock \citep[e.g.][]{pohl}. This fact led \citet{atoyantycho} to build a two--zone model for SNRs in which particles are accelerated in a small region around the shock wave (zone 1), where the magnetic field is amplified. Particles are subsequently transported (through advection and diffusion) further inside the SNR (zone 2), where the magnetic field strength may be smaller. In Atoyan \& Dermer's model, the acceleration region is much smaller than the inner region, and thus the electrons spend most of the time in the latter. Interestingly, this allowed them to decouple the region where electrons are accelerated (zone 1, where the magnetic field strength is large) from the one in which they suffer most of the synchrotron losses (zone 2, where the field strength is smaller). Assuming the existence of two zones  with different magnetic field can significantly affect the predictions of the electron spectrum and thus of  the leptonic gamma--ray emission from SNRs. Here we adopt the following simplified view: we neglect particle diffusion and we assume that electrons, after being accelerated in zone 1 are quickly advected into zone 2, which is characterized by a low field. In this case the maximum energy of accelerated electrons is computed through Eq.~\ref{eq:Emaxe}, where $B_{down}$ is the amplified field, while in order to compute the energy of the break in the spectrum (see Eq.~\ref{eq:tausyn} and following discussion) we adopt a smaller value for the magnetic field. As an illustrative example, we adopt here a constant value of 30 $\mu$G for the field strength in the inner region (the effect of changing this field will be discussed in Section~\ref{sec:discussion}). 

\begin{figure}
\includegraphics[width=.5\textwidth]{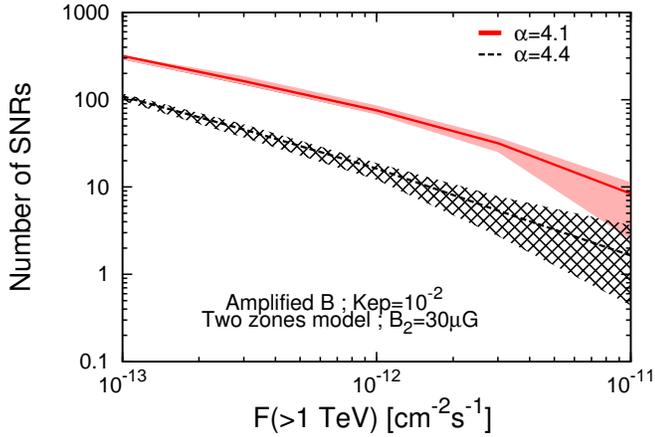}
\caption{Same as in Fig.~\ref{fig:nonamplified} and \ref{fig:amplified} but now two zones are considered: the acceleration zone around the shock, where the magnetic field is amplified and given by Eq.~\ref{eq:Bdown}, and an inner region with a weaker magnetic field equal to 30 $\mu$G. The spectrum of particles accelerated at the shock is a power law in momentum with slope 4.4 and 4.1 for the black and red curve, respectively (models M5 and M6 in Table~\ref{tab:models}).}
\label{fig:twozones}
\end{figure}

Results for this two--zone model are shown in Fig.~\ref{fig:twozones}. The black curve has been computed for a CR spectrum at injection with slope $\alpha = 4.4$ and for $K_{ep} = 10^{-2}$. The magnetic field at the shock is the amplified one (Eq.~\ref{eq:Bdown}) while the field in zone 2 is $30~\mu$G. Due to the lower value of the magnetic field in zone 2, synchrotron losses are less severe and the break in the electron spectrum moves upward in energy, thus enhancing the inverse Compton emission. This explains the larger number of gamma--ray bright SNRs expected in this case, when compared to the one zone model illustrated in Fig.~\ref{fig:amplified}. The number of SNRs with integral gamma--ray flux above 1\% of the Crab flux is $\approx 57$ while the mean value for the expected number of very bright SNRs with integral flux above $F(>1~{\rm TeV}) = 10^{-11}$cm$^{-2}$s$^{-1}$ is $\approx 1.6$.
 
Finally, the red curve in Fig.~\ref{fig:twozones} shows the expectations for a hard spectrum of the accelerated CRs with $\alpha = 4.1$ (all the other parameters are left unchanged). The evident effect of an hard spectrum is a large increase of the number of gamma--ray SNRs. In this case, the expected number of SNRs with integral flux above 1\% of the Crab is unreasonably large $\approx 190$, while the mean number of very bright SNRs with flux above $10^{-11}$cm$^{-2}$s$^{-1}$ is $\approx 8.1$, also exceedingly large. This clearly disfavor a scenario in which SNRs accelerate hard spectra of particles. Spectra significantly steeper than $\alpha = 4$ are needed to be consistent with observations, if a standard $\approx 10\%$ CR acceleration efficiency is assumed. This point will be further discussed in the next Section.

Finally, it is instructive to estimate the total number of SNRs which are currently in the Sedov stage of their dynamical evolution. This would provide a strict (and clearly over--optimistic) upper limit for the number of possible detections in gamma--rays, since CR production is believed to be efficient during this phase of the SNR evolution. By assuming a duration of the Sedov phase equal to a few $10^4$ yr and 3 supernova explosions per century in the Galaxy, this number turns out to be $\approx 1000$. Thus, for the cases considered above, even for the most optimistic, the SNRs with TeV gamma--ray fluxes above the level of 1\% of the Crab are a small fraction ($\approx$~0.01...0.1) of the total number of SNR which are believed to accelerate CRs in the Galaxy. 
 
For the reader's convenience, the parameters which have been used to compute the curves in Figures~\ref{fig:nonamplified}, \ref{fig:amplified}, and \ref{fig:twozones} are listed in Table~\ref{tab:models}. 
 
 \begin{table}
\centering
\begin{tabular}{c c c c c}
\hline 
Model & ~~~~$\alpha$~~~~ & ~~~~$K_{ep}$~~~~ & amplified B & \# of zones \\
\hline 
\hline
M1 & 4.4 & $10^{-5}$  & OFF & 1 \\
M2 & 4.4 & $10^{-2}$ & OFF & 1   \\
M3 & 4.4 & $10^{-5}$ & ON & 1  \\
M4 & 4.4 & $10^{-2}$ & ON & 1  \\
M5 & 4.4 & $10^{-2}$ & ON & 2 \\
M6 & 4.1 & $10^{-2}$ & ON & 2 \\
\hline
\end{tabular}
\caption{Values of the parameters adopted to compute the curves in Figures~\ref{fig:nonamplified} (model M1 and M2), \ref{fig:amplified} (M3 and M4), and \ref{fig:twozones} (M5 and M6). $\alpha$ is the slope of the spectrum of CRs accelerated at the shock, and $K_{ep}$ is the electron--to--proton ratio. The last two columns specify whether or not magnetic field amplification has been taken into account, and the number of zones adopted to compute the inverse Compton radiation from electrons (see text for more details).}
\label{tab:models}
\end{table}

\section{Supernova remnants and sky surveys in the TeV energy domain}
\label{sec:hess} 

In this section we perform a comparison between the predictions described above and the data currently available in the TeV gamma-ray domain. With this respect, the data obtained by the H.E.S.S. array of Cherenkov telescopes seem to be the most appropriate. Due to the large instrumental field of view ($\approx 5^{\circ}$) it has been possible to devote a significant fraction of the total observing time of H.E.S.S. to a scan of the galactic plane. The results of this scan have been published in a series of papers \citep{scan1,scan2,scan3}. The aim of the scan is to obtain a good compromise between the fraction of the sky covered by the survey and the depth and spatial homogeneity of the exposure. 
The original H.E.S.S. survey covered the range of $|l| < 30^{\circ}$ in galactic longitude and $|b| < 3^{\circ}$ in latitude \citep{scan1}, and it has been gradually extended thereafter, especially in longitude. 
To date, an extension in the range of $l = 250$ to $65$ degrees was reported \citep{scan3}. However, from Fig.~2 in that paper, it can be noticed that the exposure, and thus the sensitivity within the survey region is non-uniform. For this reason, in the following we restrict our attention to the region of galactic longitude $|l| < 40$ degrees only, within which the sensitivity for point sources is quite homogeneous and always at the level of at least $\approx 1.5\%$ of the Crab level (i.e. $F(> 1~{\rm TeV}) \approx 3.4 \times 10^{-13}$cm$^{-2}$s$^{-1}$).

\begin{table*}
\centering
\begin{tabular}{c c c c c c}
\hline 
Name & ~~~~$F(>1~{\rm TeV})$~~~~ & ~~~~d~~~~ & ~~~~age~~~~ & ~~~~radius~~~~ & ~~~~Ref.~~~~ \\
&  $[10^{-12}{\rm cm}^{-2}{\rm s}^{-1}]$ & $[{\rm kpc}]$ & $[{\rm kyr}]$ & $[^{\circ}]$ & \\
\hline 
\hline
RX J1713.7--3946 & 15.5 & 1 & 1.6 & 0.65 & 1,2,3 \\
HESS J1731--347 & 6.9 & 2.4...4 & 27 & 0.25 & 4,5 \\
CTB 37B & 0.4 & 13.2 & 0.3...3 & 0.03 & 6,7 \\
\hline
\end{tabular}
\caption{Gamma--ray fluxes, distances, ages and apparent sizes of the three SNR shells detected by H.E.S.S. in the region $|l| < 40^{\circ}$, $|b| < 3.5^{\circ}$ at a flux level above 1.5\% of the Crab. {\it References:} 1) \citealt{hessRXJ}; 2) \citealt{yoshi}; 3) \citealt{wang}; 4) \citealt{J1731}; 5) \citealt{tian}; 6) \citealt{CTB37B}; 7) \citealt{nakamura}}
\label{tab:shells}
\end{table*}

The number of TeV gamma--ray sources in the H.E.S.S. Source Catalog\footnote{\url{www.mpi-hd.mpg.de/hfm/HESS/pages/home/sources/}} within the region we selected ($|l| < 40^{\circ}$, $|b| < 3^{\circ}$) and with a flux above 1.5\% of the Crab is 35. Notably, three of them,  RX~J1713.7--3946, HESS~J1731--347, and CTB 37B, are associated with SNR shells. The physical properties of these three sources are listed in Tabel~\ref{tab:shells}. In addition to that, other three sources, CTB~37A, HESS~J1745-303, and W28, are or might be associated with SNR shells in interaction with massive molecular clouds. However, the gamma--ray emission from these interacting systems might have a different origin than the one we investigate here. For example, the gamma--ray emission from the old SNR W28 (the estimated age is a few times $10^4$~yr) has been interpreted as the results of the interactions of CRs that escaped the SNR and penetrated into the molecular cloud \citep{gabiciw28,lara}. Also the SNR coincident with the gamma--ray source HESS~J1745-303 is believed to be quite old (more than $\approx 10^4$ yr, \citealt{hess1745}), and thus also in this case an interpretation of the gamma--ray emission in terms of escaping CRs \citep{gabici09} or of re--acceleration of pre-existing CRs \citep[the so called cloud crushed model,][]{yas} might be preferred. The situation is different for the SNR CTB~37A, which is likely to be young ($\approx 1...3 \times 10^3$ yr, \citealt{CTB37A}) and thus in this case the gamma--ray emission might be related to the ongoing acceleration of CRs at the SNR shock, though the presence of the cloud might significantly affect the general picture of diffusive shock acceleration described in Sec.~\ref{sec:crgamma}.
However, the gamma--ray emission from CTB 37A can also be attributed to a pulsar wind nebula present in the region. For these reasons we do not consider these three objects in the following.
Finally, 17 out of the 35 TeV sources detected by H.E.S.S. in the region in exam still remain unidentified\footnote{The classification of TeV sources in shells, shells interacting with molecular clouds, and unidentified sources has been established by cross--correlating the H.E.S.S. source catalogue with the TeVCat online catalogue, maintained by S. Wakely and D. Horan \url{http://tevcat.uchicago.edu/}}. 
Thus, we can conclude that the number of SNRs detected in TeV gamma rays in the considered region spans from a pessimistic tally of 3 (if only the three isolated shells listed in Table~\ref{tab:shells} are considered) to an overoptimistic one of $\sim 20$ (in the unlikely event that most or all of the unidentified H.E.S.S. sources are indeed SNRs).

\begin{figure}
\includegraphics[width=.5\textwidth, height=10cm]{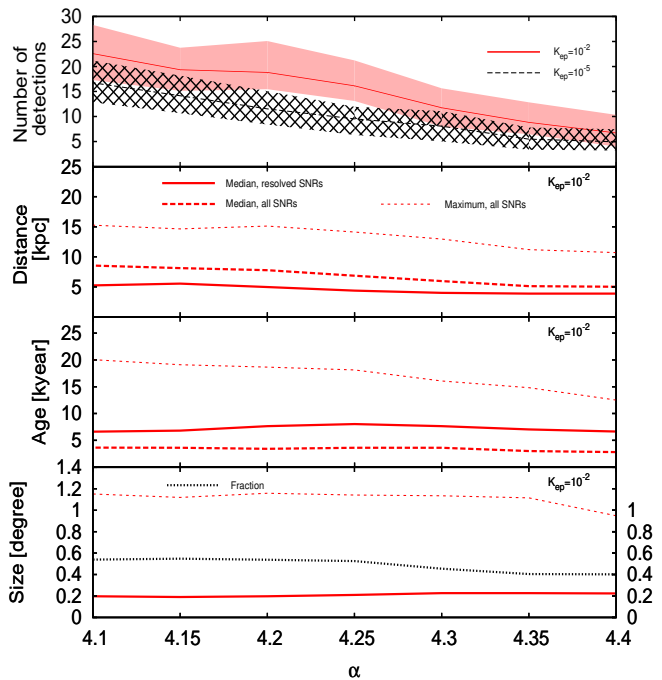}
\caption{The top panel shows the expected number of SNRs detectable by H.E.S.S. in the region of coordinates $|l| < 40^{\circ}$ and $|b| < 3^{\circ}$ as a function of the spectral slope $\alpha$ of accelerated particles. Black and red lines correspond to $K_{ep} = 10^{-5}$ and $10^{-2}$, respectively. The other panels (top to bottom) show, for the case $K_{ep} = 10^{-2}$ the distance, age, and angular size of the detected SNRs. Median values of these quantities are shown for all SNRs and for spatially resolved ones with a solid and dashed thick line, respectively. The thin dashed lines represent the maximum value for these quantities (averaged over the Monte Carlo realizations). In the bottom panel the fraction of point--like sources is also shown as a black dotted line.}
\label{fig:multiplot}
\end{figure}

To compare these numbers with our predictions we run 1000 Monte Carlo realizations of the supernova explosions in the Galaxy and compute the number of sources expected to be detected by H.E.S.S. within the region in exam. A sensitivity at the level of 1.5\% of the Crab flux has been adopted for point like sources, while for extended ones the sensitivity has been degraded by a factor of $\vartheta_s/\vartheta_{PSF}$, where $\vartheta_s$ is the source apparent size and $\vartheta_{PSF} \approx 0.1^{\circ}$ is the angular resolution of H.E.S.S\footnote{A discussion of the procedure to determine the extension of a TeV source clearly goes beyond the scope of this paper. However, it is important to remind that the classification of a sources as extended may depend on several factors, including the available photon statistics. The value $0.1^{\circ}$ adopted here must be considered as an indicative figure only.}. The results of this computation are shown in the top panel of Fig.~\ref{fig:multiplot}, where the mean number of expected detections is plotted as a function of the spectral slope $\alpha$ of accelerated CRs. The black and red lines refer to values of the electron--to--proton ratio of $K_{ep} = 10^{-5}$ and $10^{-2}$, respectively. A two--zone model has been adopted to describe the SNR, with a small region around the shock where the magnetic field is strongly amplified (see Eq.~\ref{eq:amplifiedB}), and an inner region inside the SNR where the magnetic field is significantly lower due to damping. We assume a field intensity of $30 \mu$G in the inner region, but we discuss in the next section which effect a change in the value of this parameter has.

It can be seen from Fig.~\ref{fig:multiplot} that for steep particle spectra ($\alpha = 4.4$) the expected number of detections is $\approx 5$ and 7 for $K_{ep} = 10^{-5}$ and $10^{-2}$, respectively. These numbers progressively increase if harder and harder spectra are considered, and for $\alpha = 4.1$ we found, for the two values of $K_{ep}$, $\approx$ 17 and 22 detections. These numbers are comparable, or even larger than the most optimistic estimate for the actual number of SNRs detected by H.E.S.S. in the region. This fact suggests that steep spectra are consistent with observations, while hard spectra, with $\alpha$ close to 4 (the standard prediction of test--particle first order Fermi acceleration) would imply an unreasonably high number of detections.
This finding is in agreement with both gamma--ray observations of some individual SNRs like Tycho \citep[e.g.][]{fermitycho,giovannitycho} and recent theoretical developments \citep[e.g.][]{vladimirdrift,bellsteep,damiano} that seem to indicate that diffusive shock acceleration might produce spectra significantly steeper than $\alpha = 4$ if diffusive shock acceleration is treated in a fully self--consistent manner.

All the other panels of Fig.~\ref{fig:multiplot} refer to the case $K_{ep} = 10^{-2}$, and show the median and maximum values of the following quantities (top to bottom): distance, age, and apparent angular size of the SNRs that should have been detected by H.E.S.S. in the region we are considering. Thick red lines show the median values for all those SNRs (dashed lines) and for spatially resolved ones (solid lines), i.e., with an angular size larger than $\approx 0.1^{\circ}$. The thin dashed lines show the maximum value for these quantities, averaged over the number of Monte Carlo realizations.

The median distance of detected SNRs lays, for all the values of the spectral slope $\alpha$, in the range $5~{\rm kpc} \lesssim d < 10~{\rm kpc}$, which means that in the majority of the cases the detected SNRs are closer than the galactic centre. The median distance is slightly smaller ($d \approx 5~{\rm kpc}$) if only resolved sources are considered, as expected given the worse instrument sensitivity in detecting extended sources, and given that it is easier to resolve nearby sources. On the other hand, the maximum distance up to which SNRs are detected -- a sort of horizon for the detection of SNRs -- is $\approx 15~{\rm kpc}$ for hard particle spectra ($\alpha \approx 4$) and decreases gradually for steeper and steeper spectra reaching a value of $\approx 10~{\rm kpc}$ for $\alpha = 4.4$.

The predicted median age of the SNRs detectable by H.E.S.S. is quite insensitive to the value of $\alpha$, and is of the order of $\lesssim 5~{\rm kyr}$. This slightly increase to $\gtrsim 5~{\rm kyr}$ if only resolved SNRs are considered. This is expected, given that older SNRs are obviously larger than younger ones. Also in this case, the maximum age of the detected SNRs is predicted to decrease from $\approx 20~{\rm kyr}$ to $\approx 12~{\rm kyr}$ when the spectral slope of accelerated CRs goes from $\alpha = 4.1$ to 4.4.

Finally, the predicted fraction of point--like SNRs is in the range $\approx 0.4 ... 0.6$, as indicated by the black dotted line in the bottom panel of Fig.~\ref{fig:multiplot}. Amongst extended sources, the expected median angular size is $\approx 0.2^{\circ}$, while the largest detectable sources have a size of $\approx 1 ... 1.2^{\circ}$. All these quantities are quite insensitive to the value chosen for $\alpha$.

Though a rigorous comparison between our predictions and available data is not easy, it is evident that a qualitative agreement between data and predictions exists. Our expectations for the selected region of the H.E.S.S. scan seem to reproduce correctly the actual number of detections and the typical distances, ages, and apparent sizes of the gamma--ray bright SNRs (see e.g. Fig.~\ref{fig:multiplot} and Table~\ref{tab:shells}). Moreover, as already discussed in Sec.~\ref{sec:montecarlo}, also the number of very bright (flux at the level of $\approx 10^{-11}~{\rm cm^{-2} s^{-1}}$) SNRs detectable in the whole Galaxy seems to be well reproduced. All these facts are encouraging and provide additional support to the consistency of the SNR paradigm for the origin of CRs. A summary of the main finding for the different scenarios considered in this paper can be found in Table~\ref{tab:numbers}.

\begin{table*}
\centering
\begin{tabular}{l c c c c c c}
\hline 
Model: & M1 & M2 & M3 & M4 & M5 & M6 \\
\hline 
Mean (median) number of detections: & $0.9~(2)$ &  $1.8~(3)$ &  $5.3~(6)$ & $5.9~(6)$ & $6.6~(7)$ & $22~(23)$ \\
Median distance [kpc]: & 2.6 & 2.7 & 5.0 & 5.3 & 5.0 & 8.7 \\
Median age [kyr]: & 1.8 & 1.0 & 4.2 & 3.0 & 2.8 & 4.2 \\
Median apparent size [$^{\circ}$]*: & 0.25 & 0.28 & 0.22 & 0.26 & 0.22 & 0.20 \\
Fraction of point sources: & 0.34 & 0.41 & 0.40 & 0.51 & 0.40 & 0.55 \\
Fraction of hadronic sources: & 1 & 0.59 & 1 & 0.98 & 0.87 & 0.71\\
\hline
\scriptsize{* Extended sources only (i.e. size larger than 0.1$^{\circ}$).}
\end{tabular}
\caption{Expected number of detections in the region considered in Fig.~\ref{fig:multiplot} and SNR properties for the different models listed in Table~\ref{tab:models}.}
\label{tab:numbers}
\end{table*}

We conclude the Section with two more predictions of our calculations. The first one concerns the fraction of gamma--ray bright SNRs whose emission is dominated by hadronic processes. This fraction, as can be seen from Table~\ref{tab:numbers}, depends quite strongly on the adopted parameters (especially on the magnetic field strength). It can range from $\approx 60\%$ to 100\%. Determining the hadronic or leptonic origin of the gamma--ray emission from a given SNR is a very difficult task. Consider, for example, the three SNR listed in Table~\ref{tab:shells}. While multi--wavelength observations of RX J1713.7--3946 seem to point towards a leptonic origin of the gamma--ray emission (but see \citealt{fukui} for an alternative explanation), for the other two SNRs the situation is still ambiguous. Thus, if the commonly accepted interpretation of the observations of RX~J1713.7--3946 is correct,  at least for some SNRs the detected gamma--ray emission must be leptonic, and this would disfavor models M1 and M3, for which the electron--to--proton ratio is very small ($K_{ep} = 10^{-5}$), and also model M4 for which a very large magnetic field has been assumed. 
The second prediction is the fact that a very large fraction ($\approx 80\%$ for model M5, $\approx 65\%$ for model M2) of the SNRs which are expected to be detected in TeV gamma--rays are of type Ia. The difficulty of detecting core--collapse supernovae is connected to the fact that the SNR shock propagates in the tenuous medium of the wind--blown bubble, which strongly reduces the gamma--ray emission due to proton--proton interactions (the core--collapse SNRs which are expected to be detected in gamma--rays are characterized by a dominant leptonic emission). Determining the type of the progenitor supernova is a very difficult task. Amongst TeV--bright SNRs, only a few have been firmly identified as thermonuclear or core--collapse supernovae. From the detection of the light echoes of the supernova explosions, Tycho has been firmly identified as a type Ia supernova \citep{krausetycho}, while Cas A as a type IIb \citep{krausecasa}. Also SN~1006 is confidently identified as a type Ia supernova due to its location quite distant from the galactic disk \citep[e.g.][]{1006}. For the other TeV SNRs the situation is less clear, though some hints have been provided in some cases \citep[see e.g.][]{tian2}, and thus more efforts are needed in order to increase the number of firm identifications of the supernova type.

\section{Discussion of the results}
\label{sec:discussion}

In this Section we investigate how the results obtained in this paper change when different assumptions are made on the values of some key physical parameters.

One of the most important quantities involved in our calculations is the global CR luminosity of the Galaxy, $L_{CR}^{MW}$, which represents the total energy output from all the sources of CRs in the Galaxy. If SNRs are the main sources of CRs, then from the supernova rate in the Galaxy $\nu_{SN}$ it is possible to constrain the typical amount of energy that each SNR must convert into CRs (see Eq.~\ref{eq:efficiency}). The CR luminosity of the Galaxy is determined by modeling the escape of CRs from the Galaxy, and different approaches lead to very similar values of this quantity, which is of the order of $L_{CR}^{MW} \approx 10^{41}$~erg/s \citep[e.g.][]{dogiel,CRMW}. However, an uncertainty up to a factor of $\approx$~2 might still be accepted. For this reason, we repeat our calculations for three values of $L_{CR}^{MW}$, namely, $5 \times 10^{40}$~erg/s, $10^{41}$~erg/s, and $2 \times 10^{41}$~erg/s, and, as done in Sec.~\ref{sec:hess}, we compute the expected mean number of SNRs detectable within the H.E.S.S. survey of the galactic disk in the galactic longitude and latitude ranges $|l| < 40^{\circ}$ and $|b| < 3^{\circ}$, and with integral flux above 1.5\% of the Crab. For the three values of $L_{CR}^{MW}$ discussed above the mean number of detections scales roughly linearly and reads 3.1, 6.6, and 11, respectively (for model M5 in Table~\ref{tab:models}).
The approximate linearity between the average number of detections and the CR power in the Galaxy can be easily understood as follows: if we keep all the other parameters unchanged, the effect of varying the global CR luminosity is reflected into a different acceleration efficiency $\eta_{CR}$ that the SNRs must achieve in order to sustain the CR intensity in the Galaxy. If  $L_{CR}^{MW}$ is increased by a factor of $f$, then also the acceleration efficiency $\eta_{CR}$ is augmented by the same factor, as well as the expected gamma--ray luminosity from each SNR. This in turn implies that SNRs would be visible by the same telescope up to distances $d$ a factor of $\sim f^{1/2}$ larger, or, if as a first approximation we assume SNR to be homogeneously distributed in a flat disk, within a volume a factor of $\propto d^2 \propto f$ larger, which explains the linearity. 

Note that an identical linear scaling has to be expected also if  we relax the assumption of equality $\eta_{CR} \approx \xi_{CR}$ between the two CR acceleration efficiencies defined in Sec.~\ref{sec:crgamma} and we substitute it with the expression $\eta_{CR} = f ~ \xi_{CR}$, where $f$ accounts for possible deviations from the equality. However, as already said above, theoretical investigations indicates that $f$ should be quite close to 1 \citep{damiano,vladimir2012}, and thus the predictions reported here can be regarded as fiducial estimates, easy to be rescaled for possibly different values of $f$.

Another crucial parameter is the supernova explosion rate in the Galaxy, $\nu_{SN}$. We adopt throughout the paper a value of $\nu_{SN} = 3$/century, in agreement with recent estimates \citep[e.g.][]{SNrate}. However, also in this case a systematic uncertainty of a factor of $\sim 2$ is expected \citep{SNrate}. If we repeat the estimate of the mean number of detections (model M5, $|l| < 40^{\circ}$) for explosion rates in the range $\nu_{SN} = 1 ... 3$/century we do not obtain any significant difference in the predicted number of detections. This can be understood by noting that a change in the supernova rate also affects the CR acceleration efficiency per SNR. If the rate of supernova explosions $\nu_{SN}$ is multiplied by a factor of $f$, the acceleration efficiency per SNR (and thus its gamma--ray emission) has to be multiplied by the inverse factor $f^{-1}$, in order to keep the CR luminosity in the Galaxy constant. In other words, there will be $f$ more SNRs that contribute to the CR intensity in the Galaxy, but each SNR will be a factor of $f$ less powerful in gamma--rays, and thus visible within a volume a factor of $f$ smaller. And this explains why our predictions are quite insensitive to the choice of the parameter $\nu_{SN}$. This can be restated in another way: if we reduce $\nu_{SN}$, the fraction of SNRs that can be detected by a given instrument is larger, even if the total number of detections is unchanged. 

Also the value of the magnetic field plays a crucial role and it is thus mandatory to investigate how our predictions change if different values of the field are assumed.  A smaller magnetic field strength reduces the synchrotron losses of electrons that can thus radiate more gamma--ray photons through inverse Compton scattering. We considered the two--zone model (M5) and varied the intensity of the field in zone 2. The mean number of expected detections in this case goes from $\approx 8.6$ to $\approx 6$ if the field is assumed to vary from 5 to 40 $\mu$G. Moreover, the fraction of SNRs whose TeV emission is dominated by the hadronic component goes from 63\% to 97\%.

We investigate now which effect has on our predictions a different assumption concerning the maximum energy up to which particles can be accelerated at SNR shocks. The estimates for the maximum energy given in Eq.~\ref{eq:Emax1} and \ref{eq:Emax2} are very plausible guesses,  but it is true that we are far from having a solid knowledge of the details of the amplification mechanism of the magnetic field that determines the maximum particle energy at a shock. A way to change the value of the maximum energy of accelerated particles is to change the size of the CR shock precursor, i.e. the value of the parameter $\zeta$ in Eq.~\ref{eq:confinement}. If we go from $\zeta = 0.1$ to $\zeta = 0.05$ we reduce by a factor of 2 both the size of the precursor and the value of the maximum energy. The number of expected detections is, in the two cases, $\approx 6.6$ and $\approx 5$, respectively. Thus, we can conclude that our predictions are not much affected, unless the assumed values for the maximum particle energy are varied significantly (more than a factor of 2). 

Finally, we checked for the stability of our predictions against variations of the spatial distribution of SNRs and gas in the Galaxy. We repeated the procedure for a spatial distribution of SNRs with and without taking into account the presence of spiral arms, and we used the radial distribution of SNRs given by \citet{case} instead of the one by \citet{lorimer} and did not found any significance variation in our predictions. This is connected to the fact that SNRs can be detected up to quite large distances, of the order of $\approx 10...15$~kpc (see Fig.~\ref{fig:multiplot}), and thus the effects of different spatial distributions are not that important.
We also used cylindrical symmetric templates for the gas distribution in the Galaxy (as in \citealt{shibata}), without finding an appreciable effect. However, it has to be noted that the surveys of CO and HI that are used to infer the gas distribution in the Galaxy are characterized by a spatial resolution along the line of sight of $\approx 50...100$~pc. This might create problems, for examples, in identifying dense molecular clouds which have a typical size well below 100 pc. Though this is expected to have an effect on our estimates, we know from observations that SNRs in interaction with molecular clouds are generally quite old, with ages of the order of $\approx 10^4$~yr or more \cite[e.g.][]{yas}, while the majority  of the SNRs for which we predict a detectable TeV emission have an age well below 5 kyr (see Fig.~\ref{fig:multiplot}). It is not clear whether such old SNRs are capable of accelerating particles up to energies well above $\approx 10$~TeV, as needed in order to produce $\approx$~TeV photons. According to our current knowledge of the shock acceleration mechanism, which we briefly reviewed in Sec.~\ref{sec:crgamma}, it seems that old SNRs can, at best, marginally reach these energies. This suggests that, with this respect, our predictions can still be considered solid and reliable.

\section{Conclusions}
\label{sec:conclusions} 
 
In this paper we performed a comparison between the expectations of the SNR paradigm for the origin of galactic CRs and the available data in the TeV energy domain.
Instead of proceeding on a case--by--case study of individual SNRs, we aimed at studying TeV--bright SNRs as a population. To our knowledge, this is the first time that such an approach is performed.

We started by assuming that SNRs are the main sources of CRs, and this allowed us to estimate the typical CR acceleration efficiency per SNR.
We then used a Monte Carlo approach to simulate the position and time of explosions of the SNRs in the whole Galaxy and we computed then the expected number of SNRs that should be detected in the TeV domain by the present generation of Cherenkov telescopes. To compare our predictions with data, we selected a region of the galactic disk with galactic longitude $|l| < 40^{\circ}$, for which H.E.S.S. performed a scan with a roughly spatially homogeneous exposure, corresponding to a sensitivity of $\approx 1.5\%$ of the Crab.
Predictions seem to agree well with available data, thus providing an additional consistency check of the idea that SNRs are the sources of CRs.

Our main findings can be summarized as follows: first of all, we obtained evidence for the fact that particle spectra significantly steeper than $\alpha = 4$ have to be accelerated at SNRs, if they indeed are the sources of CRs. The reason for that is the fact that hard spectra ($\alpha \approx 4$) would result in a very strong TeV luminosity and this would be inconsistent with the scarce number of SNRs currently detected at TeV energies. Secondly, the expected fraction of gamma--ray bright SNRs whose emission is dominated by neutral pion decay strongly depends on the assumptions made on the strength of the magnetic field. For the range of parameters investigated in Sec.~\ref{sec:hess} this fraction spans from $\approx 60\%$ to 100\%, and the largest values are obtained either for a very low electron--to--proton ratio or for a very large magnetic field strength (one of the two conditions suffices to increase the fraction up to $\approx 100\%$). The fact that there is at least one SNR (namely RX~J1713.7--3946) whose gamma--ray emission is commonly ascribed to inverse Compton scattering might suggest that an high electron--to--proton ratio (of the order of $K_{ep} \approx 10^{-2}$) characterizes the acceleration of particles at SNRs and that regions where the field strength is not too large must exist in SNRs. Finally, according to our predictions we should expect to detect the same number of extended and point--like (where with point--like we intend sources with a size smaller than 0.1 degree) TeV--bright SNRs, and supernovae of type Ia should account for a large fraction of the detections ($\approx 60 ... 80\%$).

Before concluding we comment on a possible extension of the procedure described in this paper to the GeV energy domain, currently probed by the Fermi and Agile satellites.
Remarkably, a constantly increasing number of SNRs is being detected in the GeV energy band. Several of these SNRs are quite old, often radiative systems that show clear signatures of interactions with massive molecular clouds \citep{yas}. For these systems, not considered here, the scenario of particle acceleration and gamma--ray production is most likely very different from the one considered in this paper \citep[see e.g.][]{gabici09, malkov, yas}. 
Despite this fact, we can anyway use the procedure developed in this paper, and keep in mind that the estimates obtained in this way in the GeV domain would be very approximate, and most likely lower limits only (because it won't take into account the old interacting systems). By considering a sensitivity for Fermi (integrated above 1 GeV) of few $10^{-9} {\rm cm^{-2} s^{-1}}$ in the inner Galaxy (where most of the detections are likely to happen) we obtain a number of expected detections of the order of several tens.

In a forthcoming publication (Cristofari et al., in preparation) the procedure developed in this paper will be used to estimate the impact that the next generation Cherenkov Telescope Array will have on the studies of acceleration of CRs at SNR shocks.


\section*{Acknowledgments}

We thank E. de O\~na Wilhelmi, A. Djannati-Atai, F. Giordano, D. Maurin, G. Morlino, L. Nava, and M. Renaud for helpful discussions. SG, RT, and EP acknowledge support from ANR under the JCJC Programme. SG acknowledges support from the EU [FP7 - grant agr. n$^o$256464]. SC acknowledges support from the city of Paris under the programma "Research in Paris". RT acknowledges support under the Labex UnivearthS.

\label{lastpage}

\end{document}